\documentstyle[prl,aps,twocolumn,epsf]{revtex}

\begin{document}

\pagestyle{myheadings}
\markboth{Hiroki Saito and Masahito Ueda}{Hiroki Saito and Masahito Ueda}

\draft

\title{Quantum-Controlled Few-Photon State Generated by Squeezed Atoms}

\author{Hiroki Saito${}^1$ and Masahito Ueda${}^{1,2}$}
\address{${}^1$Department of Physical Electronics,
Hiroshima University, Higashi-Hiroshima 739, Japan}
\address{${}^2$Department of Physics, University of Illinois at
Urbana-Champaign, Urbana, IL61801-3080}

\date{\today}

\maketitle

\begin{abstract}
General principles and experimental schemes for generating a desired 
few-photon state from an aggregate of squeezed atoms are presented.
Quantum-statistical information of the collective atomic dipole
is found to be faithfully transferred to the photon state even in a 
few-photon regime.
The controllability of few-photon states is shown to increase
with increasing the number of squeezed atoms.
\end{abstract}

\pacs{42.50.Dv, 03.65.Bz, 42.50.Gy, 42.50.Lc}

\narrowtext

One of the main aims in quantum optics has been to manipulate 
quantum-statistical properties of the electromagnetic field. 
Since the first milestone of generating squeezed state of light 
was successfully achieved~\cite{squeeze}, considerable efforts have
been devoted towards the production of number state whose 
average photon number is less than a few tens~\cite{Rempe,Yamanishi}.
If the average photon number is much greater than this, the necessity
of using nonclassical light virtually disappears because the coherent 
state having a few tens of photons already has a sufficiently 
low bit error rate required for optical communication and precision 
measurement. 
Photons also carry information about the phase whose quantum
fluctuations limit the interferometric sensitivity~\cite{Kitagawa}.
In contrast to the case of photon number, methods of regulating
the phase of few-photon states have yet to be explored.
In this Letter we, for the first time, present general principles and
experimental schemes for generating a desired few-photon state.
By this method one can control not only the average and variance in 
photon number, but also the width and orientation of the uncertainty 
ellipse in phase space in any desired direction.
We will show that quantum-statistical information of the collective
atom dipole is rather faithfully transferred to those of emitted photons, 
and discuss how to exploit this property to produce a desired
few-photon state.
We can thus generate any desired few-photon state by preparing the 
atoms in some prescribed quantum state.

\begin{figure}
\begin{center}
\leavevmode\epsfysize=130mm \epsfbox{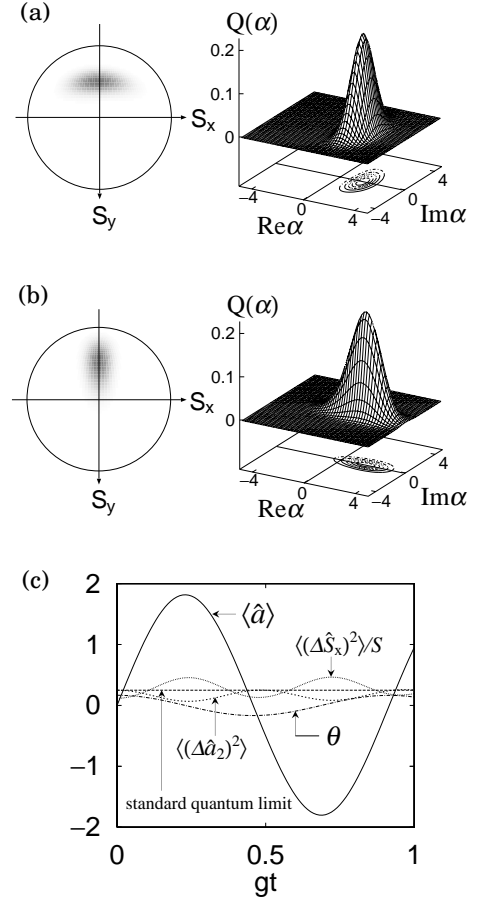}
\caption{
(a)(b): Quasi-probability distributions of squeezed fifty atoms 
with $\langle(\Delta\hat{S}_\bot)^2\rangle=2.93$, $|\langle \hat{S}
\rangle|=24.2$, $\langle\hat{S}_x\rangle=0$, and $\theta=\pi/6$
as seen from the negative $z$ axis (left) and those of the radiation 
field emitted from the atoms (right).
In (a), the amplitude is squeezed, while in (b) the phase is squeezed.
(c): Time evolutions of amplitude $\langle \hat{a}\rangle$ and variance
$\langle (\Delta \hat{a}_2)^2\rangle$ of the radiation field and 
the normalized variance of the atomic dipoles for the case of Fig. 1(b).
The standard quantum limit shows the value of 
$\langle (\Delta \hat{a}_2)^2\rangle$ for the coherent state and 
$\theta$ denotes the angle between the mean spin vector and the negative
$z$ axis.}
\label{Fig1}
\end{center}
\end{figure}

We first discuss a general condition for a collection of atoms to
be able to generate any desired few-photon state. 
Consider a 
simple situation in which a collection of atoms are placed in a 
resonant cavity and interact with a single-mode photon field. 
It is well-known that collective properties of two-level atoms,
which are placed within the photon wavelength but not too close to
avoid direct interaction between the atoms,
can be described in terms of the collective spin operators as 
$\hat S_x = \sum_i \hat\sigma_{ix} / 2$,
$\hat S_y = \sum_i \hat\sigma_{iy} / 2$, and $\hat S_z =
\sum_i \hat\sigma_{iz} / 2$,
where $\hat\sigma_{ix}$, $\hat\sigma_{iy}$, and $\hat\sigma_{iz}$ 
denote the Pauli spin operators for the $i$th atom.
Assuming the Jaynes-Cummings interaction~\cite{JC} between
the atoms and the photon field, the total Hamiltonian is given by
\begin{equation} \label{JCH}
\hat H = \hbar\omega_a \hat S_z + \hbar \omega_f
\hat a^\dagger \hat a + \hbar g (\hat a\hat S_+ + \hat a^\dagger \hat S_-),
\end{equation}
where $\hat a^\dagger$ and $\hat a$ are the creation and annihilation
operators of the photon field, $\hat S_\pm \equiv \hat S_x \pm i\hat S_y$,
$\hbar\omega_a$ is the energy difference between the two levels
of the atoms, $\hbar\omega_f$ is an energy quantum of the photon,
and $g$ is a coupling constant.
When $\omega_f = \omega_a$, we can eliminate the noninteracting part of
the Hamiltonian $H_0 = \hbar\omega_a \hat S_z + \hbar \omega_f
\hat a^\dagger \hat a$ by working on a rotating frame
$e^{i\hat H_0 t / \hbar} |\psi\rangle$.
Since we want to manipulate the width and the orientation of the 
uncertainty ellipse in phase space in any desired direction,
it is convenient to introduce operators in the direction specified by
the azimuth angle $\phi$ as
$\hat a_\phi \equiv \frac{1}{2}(\hat ae^{-i\phi} 
+ \hat a^\dagger e^{i\phi})$
and $\hat S_\phi \equiv \frac{1}{2}(\hat S_+e^{-i\phi} 
+ \hat S_-e^{i\phi})$.
These operators obey the following equations of motion:
\begin{eqnarray}
& & \frac{d\hat a_\phi}{dt}=-g\hat S_{-\phi+\pi/2},
\label{dadt} \\
& & \frac{d\hat S_{-\phi+\pi/2}}{dt}=-2g\hat a_\phi \hat S_z, \\
& & \frac{d\hat S_z}{dt}=2g(\hat a_\phi \hat S_{-\phi+\pi/2}
+ \hat a_{\phi+\pi/2} \hat S_{-\phi}).
\label{dadt1}
\end{eqnarray}
When the atoms are irradiated by coherent light with classical intensity,
the mean field approximation is valid, and Eqs.~(\ref{dadt})-(\ref{dadt1}) 
reduce to the familiar optical Bloch equations.
Since we are interested in reducing quantum fluctuations in a few-photon
regime, we have to take into account higher-order correlations.
In particular, we are interested in the variance of $\hat a_\phi$.
Its dynamical evolution is governed by
\begin{eqnarray}
\frac{d\langle(\Delta \hat a_\phi)^2\rangle}{dt} 
& = & -2g \langle (\Delta\hat a_\phi)(\Delta\hat S_{-\phi+\pi/2})
\rangle, \label{dDadt} \\
\frac{d^2\langle(\Delta \hat a_\phi)^2\rangle}{dt^2}
& = & g^2 \Bigl( 4\langle (\Delta\hat a_\phi)(\Delta\hat a_\phi
\hat S_z) \rangle \nonumber \\
& & + 2\langle(\Delta \hat S_{-\phi+\pi/2})^2\rangle
\Bigr),
\label{dDadt2}
\end{eqnarray}
where $\Delta\hat{\cal O} \equiv \hat{\cal O} - \langle\hat{\cal O}
\rangle$ for an arbitrary operator $\hat{\cal O}$.
When the field is initially in the vacuum state, there is no initial
correlation between the atoms and the field, so the right-hand side of
Eq.~(\ref{dDadt}) vanishes at $t=0$ and the first term in the square brackets
in Eq.~(\ref{dDadt2}) becomes $4\langle (\Delta \hat a_\phi)^2
\rangle \langle \hat S_z\rangle = \langle \hat S_z\rangle$.
The second derivative in Eq.~(\ref{dDadt2}) is therefore negative if
and only if
\begin{equation} \label{sqcond}
\langle(\Delta \hat S_{-\phi+\pi/2})^2\rangle <
\frac{|\langle \hat S_z\rangle |}{2} \;\;\; \mbox{and} \;\;\;
\langle \hat S_z\rangle < 0.
\end{equation}
In this case, the fluctuations in $\hat a_\phi$ will be suppressed
to below the standard quantum limit at times much shorter than $\sim g^{-1}$.
The field displacement $\langle \hat a_\phi \rangle$
and its variance $\langle(\Delta \hat a_\phi)^2\rangle$ can be
controlled independently because their time evolutions are 
governed respectively
by $\langle \hat S_{-\phi+\pi/2}\rangle$ and
$\langle(\Delta \hat S_{-\phi+\pi/2})^2\rangle$. From
Eqs.~(\ref{dadt})-(\ref{dDadt2}) we find that the amplitude
squeezed state is obtained from the atomic state that satisfies, e.g.,
$\langle \hat S_x\rangle = 0, \langle \hat S_y\rangle \neq 0,
\langle \hat S_z\rangle < 0$, and $\langle(\Delta \hat S_y)^2\rangle
< |\langle \hat S_z\rangle | / 2$.
The phase squeezed state is obtained only by replacing
$\langle(\Delta \hat S_y)^2\rangle$ for
$\langle(\Delta \hat S_x)^2\rangle$.
The right figures in Fig. \ref{Fig1} (a) and (b) illustrate generation of 
the amplitude squeezed state and the phase squeezed state from squeezed fifty 
atoms.
The left figures show the initial squeezed atom states, prepared by the 
scheme discussed below, in the spin quasi-probability distribution defined by
$\langle\theta, \phi
|\hat\rho_{\rm atom}|\theta, \phi\rangle$, where
$|\theta, \phi\rangle \equiv e^{-i\phi\hat S_z} e^{-i\theta\hat S_y}
|S, S_z = S\rangle$ is the coherent state
of spin or angular momentum and will be referred to as
the Bloch state~\cite{Arecchi}.
The quasi-probability distribution of the photon field
$Q(\alpha) \equiv {\rm Tr_{atom}} [\langle\alpha |\hat\rho|\alpha\rangle]
 / \pi$ (right figures in Fig.~\ref{Fig1} (a) and (b))
is obtained by numerical diagonalization of the Jaynes-Cummings
Hamiltonian, where $|\alpha\rangle$ is the coherent state of
the radiation field with amplitude $\alpha$, and $\hat\rho$
is the density operator of the entire system when the maximal
squeezing is obtained. From
Fig.~\ref{Fig1} we find that the profile of $Q(\alpha)$ follows that of
$\langle\theta, \phi |\hat\rho_{\rm atom}|\theta, \phi\rangle$
projected on the $S_x$-$S_y$ plane, where $S_x$ and $S_y$ correspond
to $-{\rm Im}\;\alpha$ and $-{\rm Re}\;\alpha$.
This rather faithful transfer of quantum information from the atomic
system to the photon system holds in general, and tells us how to
prepare the collective atomic state in order to produce a desired
few-photon state.
Figure 1(c) shows the time evolutions of the atomic and field
observables for the case of Fig.~1(b).
The radiation-field amplitude $\langle \hat{a}\rangle$ 
grows as the mean spin vector tilts towards the negative z-axis
(i.e., $\theta\rightarrow 0$).
We also note that quantum fluctuations in the radiation field, 
$\langle (\Delta \hat{a}_2)^2\rangle$ decreases in time at the expense of 
increasing atomic fluctuations
$\langle(\Delta \hat S_x)^2\rangle$.

For a collection of atoms to be able to radiate a photon state that
is squeezed in any desired direction in phase space, 
which we will refer to as 
{\it tailor-made radiation}, condition (\ref{sqcond}) has
to be met for arbitrary $\phi$. 
Thus the necessary and sufficient condition for the tailor-made
radiation is given by
\begin{equation} 
\langle(\Delta \hat S_\perp^{\rm min})^2\rangle 
< \frac{|\langle\hat{\bbox{S}}\rangle|}{2},
\label{SS}
\end{equation}
where $\langle(\Delta \hat S_\perp^{\rm min})^2\rangle$ denotes
the minimum value of the variance perpendicular to the mean spin vector
$\langle\hat{\bbox{S}}\rangle$.
Note that condition (\ref{SS}) is more stringent than the condition used
in Ref. \cite{Kitagawa,Wineland}
which discussed the interferometric phase sensitivity.
A crucial observation is that phase squeezing (as in Fig. \ref{Fig1} (b))
can only be obtained by states satisfying the condition (\ref{SS}).
This is because fluctuations projected on the $S_x$-$S_y$ plane
cannot be reduced to below $\langle(\Delta \hat S_\perp^{\rm min})^2\rangle$
in the direction of $\phi$ by any rotation of the spin vector.
This is why the Bloch state which has an isotropic uncertainty distribution
with respect to the plane perpendicular to the mean spin vector
cannot radiate the phase squeezed state.
The Bloch state can, on the other hand, radiate amplitude-squeezed state
by tilting the spin vector \cite{Walls}.
Nor can a single atom be used for the tailor-made radiation,
because it has no partner to be quantum-mechanically entangled with
in order to meet condition (\ref{SS}).
It should also be noted that a popular definition of the spin squeezing
\cite{Walls,Wodkiewicz,Macomber,Barnett}
\begin{equation} \label{SS1}
\langle(\Delta \hat S_i)^2\rangle < \frac{|\langle \hat S_z
\rangle |}{2} \;\;\; (i = x \; \mbox{or} \; y),
\end{equation}
cannot be used as a criterion for the tailor-made radiation because
this condition can be met by the Bloch state whose spin vector is
tilted from the $z$ axis \cite{Heidmann}.

In order to control the degree of squeezing of photons, we must
solve the time evolution (\ref{dadt})-(\ref{dDadt2}).
Although the exact solutions is unavailable because of high
nonlinearity of these equations, we can obtain approximate solution
that becomes exact when the number of atoms is large and
$\langle\hat{S}_z \rangle \sim -S$, i.e., the spin angle from
the negative z-axis is small:
\begin{eqnarray}
\langle(\Delta\hat a_\phi)^2\rangle & = & \frac{1}{4} \cos^2\sqrt{2S_0}gt
\nonumber \\
& & + \frac{\langle(\Delta\hat S_{-\phi+\pi/2})^2\rangle_0}{2S_0}
\sin^2\sqrt{2S_0}gt 
\label{deltaa} \\
\langle(\Delta\hat S_{-\phi+\pi/2})^2\rangle
& = & \langle(\Delta\hat S_{-\phi+\pi/2})^2\rangle_0
\cos^2\sqrt{2S_0}gt \nonumber \\
& & + \frac{S_0}{2} \sin^2\sqrt{2S_0}gt
\label{deltaS}
\end{eqnarray}
where $\langle(\Delta\hat S_{-\phi+\pi/2})^2\rangle_0$ denotes the variance in
the initial spin state, and the length of the mean spin vector
$|\langle\hat{\bbox{S}} \rangle|$ is assumed to be almost constant $S_0$.
These solutions are periodic functions with period $\pi(g\sqrt{2S_0})^{-1}$.
The photon fluctuation (\ref{deltaa}) attains a minimum value
$\langle(\Delta\hat S_{-\phi+\pi/2})^2\rangle_0 / (2S_0)$ at
$t = \pi(2g\sqrt{2S_0})^{-1}$, and therefore the squeezed radiation can
be obtained if the spin satisfies the condition (\ref{sqcond})
and the degree of squeezing is proportional to that of the spin.
The expressions (\ref{deltaa}) and (\ref{deltaS}) imply that quantum
fluctuation is transferred from the field to the atoms, as can be
seen in Fig.~\ref{Fig1}(c).

Several methods have been proposed to generate squeezed
spin state: the Jaynes-Cummings interaction with the coherent
state \cite{Waka}, or with the squeezed vacuum \cite{Wineland},
and interaction of the spins through nonlinear Hamiltonians
\cite{Barnett,Kitagawa}.
We focus our discussion to the first method.
In Ref.~\cite{Waka} the Jaynes-Cummings Hamiltonian (\ref{JCH})
with the initial coherent state and the initial spin state $|S, S\rangle$
are used, while in Ref.~\cite{Wineland} the interaction Hamiltonian
$\hat H_2 = \hbar\Omega(\hat a^\dagger \hat S_+ + \hat a\hat S_-)$
with the initial spin state $|S, -S\rangle$ are used, where $\Omega$
is a coupling constant.
Since these models are mathematically equivalent \cite{Note},
we restrict our attention to the former.

\begin{figure}
\begin{center}
\leavevmode\epsfysize=70mm \epsfbox{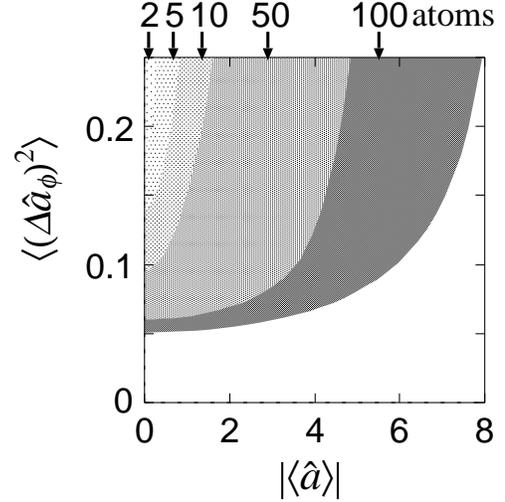}
\caption{
Possible range of amplitude $|\langle \hat{a}\rangle|$ and its 
variance $\langle (\Delta \hat{a}_\phi)^2\rangle$ of the radiation 
field that can be obtained by 2, 5, 10, 50, and 100 atoms
prepared by interaction with the coherent state.
}
\label{range}
\end{center}
\end{figure}

The quantity of our interest is the degree of squeezing of the spin
obtained by the interaction with the coherent state,
given the number of atoms.
We seek the maximally squeezed spin state
numerically, rotate it in various direction,
and use it as the initial spin state in the radiation process.
Figure \ref{range} shows the range of amplitude 
$|\langle \hat{a}\rangle|$ and
variance $\langle (\Delta \hat{a}_\phi)^2\rangle$ of the radiation 
field that can be achieved by 2 $\sim$ 100 atoms.
This shows that the larger the number of atoms,
the tunable range for the radiation field becomes wider.

\begin{figure}
\begin{center}
\leavevmode\epsfysize=70mm \epsfbox{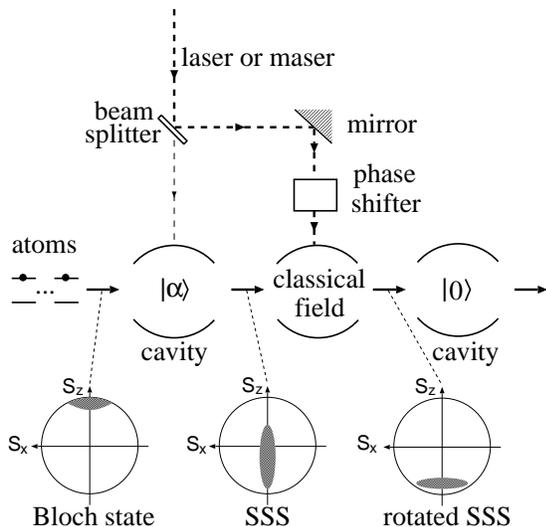}
\caption{
Schematic illustration of an experimental setup to generate a few-photon
state that features any desired quantum statistics.
The state of the atoms at each stage is shown by the spin
quasi-probability distribution.
The two-level excited atoms go into the first cavity and interact 
with a coherent state of the radiation field $|\alpha\rangle$.
The output atoms are in a squeezed spin state (SSS).
By interaction with a classical field in the second cavity, 
the mean spin vector is rotated to a desired direction, where
the rotation axis can be specified by the phase shifter.
The atoms then go into the third cavity and emit photons there.
Left in the third cavity is the desired few-photon state which can
be extracted by our changing the cavity quality factor.
}
\label{setup}
\end{center}
\end{figure}

We propose two possible experimental schemes to implement our theory.
The first one is a scheme using the micromaser technique \cite{Meschede}
as illustrated in Fig.~\ref{setup}, where the state of atoms is
indicated in the spin quasi-probability distribution at each stage.
It consists of three stages:
(1) The excited two-level atoms are injected into the
first cavity in which the atoms become squeezed by interacting with the 
coherent state of the radiation field $|\alpha\rangle$ prepared by laser 
or maser \cite{Waka}.
(2) The output squeezed atoms pass through the coherent field
with classical intensity.
This field rotates the mean spin vector in the spin space to
the desired direction.
To control the rotation axis the coherent field and the classical
field must be driven synchronously with an appropriate phase difference
provided by the phase shifter.
(3) The atoms go into the third vacuum cavity, radiate photons
and come out of the cavity before reabsorbing the emitted photons.
Left in the third cavity is the desired photon state which we can
take out by switching the Q-factor of the cavity mechanically or
by applying a magnetic field.

The second scheme is to employ the atom trapping and
the laser-cooling, in which the above three stages are
implemented at the same place. For this purpose, 
the optical cavity should be off-resonant during the preparation of
the atomic state, and be resonant only at the time of radiation.
Interaction with the coherent state corresponding to the first
stage above can be realized by interaction with the center-of-mass
oscillation of atoms through the stimulated Raman transition
\cite{Heinzen}.
This second scheme has the advantage of producing a large number of 
squeezed atoms.

In an actual experiment, we must finish the whole sequence of processes
before the two-level atoms decay into other levels.
If we use, for example, the $63{\it p}_{3/2} \rightarrow 61{\it d}_{3/2}$
transition of rubidium atoms, the lifetime is of order millisecond
and the coupling constant is $g \sim 10^4$ Hz.
Since the required interaction time is $gt \sim 1$, i.e.,
$t \sim 10^{-4}$ sec, the whole procedure can be accomplished within
the atomic lifetime.
The finite Q-factor of the cavity will not be an obstacle,
since the cavity lifetime now reaches $t_c \sim 10^{-1}$ sec
in the microwave regime~\cite{Rempe}.
Thermal photons, however, must be carefully suppressed.
If we use the above-mentioned transition ($21.5$ GHz), the temperature should
be below, say, 0.2 K in order to suppress the average number of thermal
photons in the cavity to below 0.01.

In conclusion, we have shown that the quantum-statistical information of 
collective atomic dipoles is faithfully transferred to the radiation
field even in a few-photon regime.
This implies that we can produce a desired few-photon state by preparing 
atoms in an appropriate squeezed state.
This idea can be tested using a high-Q cavity that sustains more than one
atom undergoing the Jaynes-Cummings interaction with the radiation 
field.

One of the authors (M.U.) gratefully acknowledges fruitful discussions
with F. Shimizu and M. Kuwata-Gonokami.
This work was supported by a Grant-in-Aid for Scientific
Research (Grant No. 06245103) by the Ministry of Education, Science,
Sports, and Culture of Japan, and by the Core Research for Evolutional
Science and Technology (CREST) of the Japan Science and Technology
Corporation (JST).



\begin{references}
\bibitem{Note}
 From the relation $e^{i\pi \hat S_x} \hat S_\pm e^{-i\pi \hat S_x}
= \hat S_\mp$, we find that time evolution with $\hat H_2$
can be rewritten by
\[
e^{-i\hat H_2t} |-S\rangle = e^{-i\hat H_2t} e^{-i\pi \hat S_x} |S\rangle
= e^{-i\pi \hat S_x} e^{-i\hat H_{\rm int}t} |S\rangle,
\]
which indicates that both models are equivalent rotating
the initial and final spin state.
\end{references}
\end{document}